\documentclass[photonics,article,submit,moreauthors,pdftex]{Definitions/mdpi} 
\firstpage{1} 
\pubvolume{1}
\issuenum{1}
\articlenumber{0}
\pubyear{2021}
\copyrightyear{2021}
\datereceived{} 
\dateaccepted{} 
\datepublished{} 
\hreflink{https://doi.org/} % If needed use \linebreak

\newcommand{\p}{\partial}
\newcommand{\ep}{\varepsilon}

\newcommand{\om}{\omega}
\newcommand{\nn}{\nonumber}
\newcommand{\ta}{\theta}
\newcommand{\vta}{\theta}

\newcommand{\cH}{{\cal H}}
\newcommand{\cF}{{\cal F}}
\newcommand{\cE}{{\cal E}}
\newcommand{\cW}{{\cal W}}

\newcommand{\wh}{\widehat}
\newcommand{\be}{\begin{equation}}                                       
\newcommand{\ee}{\end{equation}}
\newcommand{\ba}{\begin{eqnarray}}
\newcommand{\ea}{\end{eqnarray}}
\newcommand{\bl}{\begin{align}}
\newcommand{\el}{\end{align}}
\newcommand{\bref}[1]{(\ref{#1})}
\newcommand{\lab}[1]{\label{#1}}

\newcommand{\bsub}{\begin{subequations}}                      
	\newcommand{\esub}{\end{subequations}}     
	
\Title{Controlling Microresonator Solitons with the Counter-Propagating Pump}

\TitleCitation{Controlling Microresonator Solitons with the Counter-Propagating Pump}

\Author{Zhiwei Fan $^{1}$  and Dmitry V. Skryabin $^{1,2,}$*\orcidC{}}  %Please carefully check the accuracy of names and affiliations. The order of the author's name is different from submission system, please check

\AuthorNames{Zhiwei Fan and Dmitry V. Skryabin}

\AuthorCitation{Fan, Z.;  Skryabin, D.V.}
\address{%
	$^{1}$  \quad Department of Physics, University of Bath, Bath BA2 7AY, UK\\
	$^{2}$ \quad Russian Quantum Center, 143026 Skolkovo, Russia }

\corres{Correspondence: d.v.skryabin@bath.ac.uk }

\abstract{Considering a bidirectionally pumped ring microresonator, we provide a concise derivation of the model equations allowing us to eliminate the repetition rate terms and   reduce the nonlinear interaction between the counter-propagating waves to the power-dependent shifts of the resonance frequencies. We present the simulation results of the soliton control by swiping the frequency of the counter-propagating wave in the forward and backward directions  and with the soliton-blockade effect either present or not. We highlight the non-reciprocity of the forward and backward scans. Furthermore, we report the soliton crystals and breathers existing in the vicinity of the \mbox{blockade  interval.}
}


\begin{document}
%%%%%%%%%%%%%%%%%%%%%%%%%%%%%%%%%%%%%%%%%%
%\setcounter{section}{-1} %% Remove this when starting to work on the template.
%\section{How to Use This Template}
%	\tableofcontents
\nolinenumbers

\section{Introduction}
Frequency comb generation in the high-Q ring microresonators and the associated dissipative Kerr  solitons have reached unprecedented heights of practical relevance~\cite{rev3}. 
The complexity of the microresonator soliton properties is hard to exaggerate.  
A challenge  has emerged after a series of experiments with the bidirectionally pumped microresonators, where combs and solitons \cite{vahala,gaeta,tobias} and 
symmetry breaking \cite{chirality,pascal2} have been observed in counter-propagating waves. Studies into the gyroscope applications of these devices are also becoming increasingly 
important~\cite{vahala2,vahala3,matsko2,matsko3,ch}.

The problem of the mathematical modeling of the multimode regimes of operation of the bidirectional resonators has been addressed and to a large degree resolved only recently  by demonstrating the equivalence of the coupled-mode model derived from the ab initio  Maxwell system to  the envelope equations where the nonlinear interaction (cross-phase modulation, XPM) between the counter-propagating waves is reduced to the power-dependent shifts of the resonance frequencies and the fast oscillations with the repetition rate frequencies being eliminated~\cite{cole,osac,lobanov}. Below, we recapture the main steps of the model derivation  and 
highlight the transition from the four-envelope model accounting for the repetition rates to the two-envelope formulation that eliminates them.  

The deeper insight into the independence of the XPM nonlinearity from the phase relations between the interacting modes~\cite{cole,osac,lobanov} has recently led us to the prediction of the soliton-blockade effect~\cite{ol}. The blockade is achieved when the tuning frequency of one of the counter-rotating fields first disrupts and then restores the  soliton transmission in the other field. 
The refractive index change of the soliton field happens via the XPM effect and is particularly strong in microresonators due to  their high finesses providing 3--6 orders of magnitude boost to the circulating powers relative to the input one~\cite{bil}. This creates a variety of opportunities for efficient 
nonlinear control of optical signals (see, e.g.,~\cite{arnold,pra} and references therein), and, in particular, the soliton blockade effect   utilizes the high sensitivity of the nonlinear response of one of the fields towards the changes of the driving frequency of the counter-propagating one.

In Section \ref{sec3},  we look into how the power and frequency of the soliton and control pumps can be used to expand 
the soliton-blockade range. We also demonstrate there that the soliton-crystals, well known in the unidirectionally pumped resonators~\cite{cr1,cr2}, emerge from the edges of the blockade interval. We performed  both forward and backwards adiabatic scans of the control field frequency and demonstrate non-reciprocity of the soliton-blockade effect (see Section~\ref{sec4}). In Section \ref{sec5}, we present equations for  perturbations around the solitons, which bare the features specific to the integral nature of the XPM terms  and demonstrate  
the breather states close to the soliton-blockade interval (see, e.g.,  \cite{br1,br2,br3} for the breather studies in the unidirectional resonators).

%%%%%%%%%%%%%%%%%%%%%%%%%%%%%%%%%%%%%%%%%%

\section{Model}\label{sec2}
\subsection{Equations, Field Envelopes and Modes}
We proceed by postulating that the real electric field of a given transverse mode family  in a multi-mode ring microresonator can be expressed  as~\cite{osac} 
\be
b{\vec F}(r,z)\cE(t,\vta).
\lab{f0}
\ee

The whole of the latter expression has units of volts per meter. 
$\vta\in(0,2\pi]$ is the angular coordinate along the resonator circumference. $t$ is physical time. $b$ is the normalization constant that re-scales $\cE$ to the desirable units. $\cE^2$ has units of power in what follows.
${\vec F}(r,z)$ is the transverse mode profile normalized so that its maximum equals one. 
The spectrum of $\cE(t,\vta)$ is assumed to be sufficiently narrow to allow disregarding the dispersive changes of ${\vec F}(r,z)$. Separation of the spatial variables in Equation~\eqref{f0} is an approximation that generally works well in a typical microresonator.

$\cE$ can then be sought as a superposition of the two counter-propagating waves, 
\be	
\cE = e^{iM\vta-i\om_+ t}Q_+(\vta,t)+
	e^{iM\vta+i\om_+ t}Q_-^{*}(\vta,t)+c.c.~.
	\lab{f1}\ee
	
	Here,  $M$ is the mode number with the resonance frequency $\om_0$  and $\om_+$
	 is the frequency of the laser pumping the plus-wave, so that 
	\be
	\delta_0=\om_0-\om_+,\lab{f2}
	\ee
defines the respective detuning. 
$Q_\pm$ are the envelope functions,  which  can be expressed  via their mode expansions as
\be	Q_+ =\sum_{\mu}Q^+_\mu(t) e^{i\mu\vta},~~
	Q_- =\sum_{\mu}Q^{-}_\mu(t) e^{-i\mu\vta}.
	\lab{f3} 
	\ee
$Q_\mu^\pm$ are the time dependent mode amplitudes. $\mu=-N/2+1,\dots, 0,\dots, N/2$ are the relative mode numbers  and  $\om_\mu=\om_0+D_1\mu+\tfrac{1}{2}D_2\mu^2$
are the corresponding resonance frequencies. $D_1/2\pi$ is the resonator 
repetition rate   and $D_2/2\pi$ is dispersion. 
 
 The structure of the mode expansion in Equation~\bref{f3} assumes that
 the effects of $D_1$, $D_2$, linewidth and nonlinearity will all be embraced
 by the equations derived for $Q_\mu^\pm$ after Equations~\bref{f3} are substituted to the Maxwell equations. Using $\om_+$ in the exponents in front of the mode expansions for the clockwise and counter-clockwise fields in Equation~\bref{f3} implies that  detuning between the frequencies of the lasers pumping the \mbox{counter-rotating fields }
 \be
 \ep=\om_+-\om_- \lab{f5}
 \ee
will resurface in the equations for $Q_\mu^\pm$ (see Figure~\ref{fff0} for a schematic illustration of the \mbox{pump arrangements).}

\begin{figure}[H]
	\includegraphics[width=.98\linewidth]{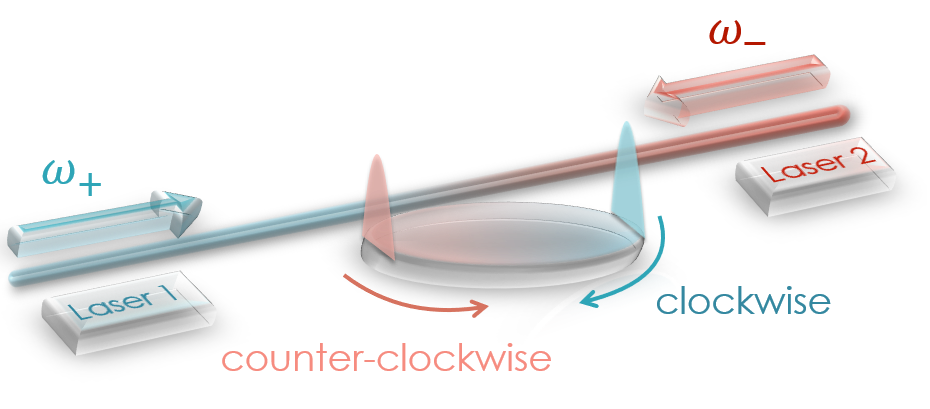}
	\caption{A schematic illustration  of the bidirectionally pumped resonator.	}
	\lab{fff0}
\end{figure}

The presence of the backscattering effects induces coupling between the counter-propagating waves. 
The authors of \cite{osac,lobanov}   formally traced what could be expected naturally, namely   the 
$Q_{\mu}^+e^{i\mu\vta}$ wave couples to the $Q_{\mu}^-e^{i\mu\vta}$ one. Comparing this coupling structure  to the definition of the $Q_-$ envelope function in Equation~\bref{f3} suggests the need to define the envelope functions for the reflected waves,
\be
Q_{+}^{(r)}=\sum_\mu Q_\mu^{+}(t)e^{- i\mu\vta},~
Q_{-}^{(r)}=\sum_\mu Q_\mu^{-}(t)e^{i\mu\vta}.
\lab{f8a}\ee
While the mode amplitudes entering the $Q_{\pm}$ and $Q_{\pm}^{(r)}$ envelopes are the same,
the signs of the exponential terms are arranged differently (cf.  Equations~\bref{f3} and \bref{f8a}). 

The envelope equations  that follow are
\begin{subequations}
	\lab{f9}
	\begin{align}
	 i\p_tQ_+&=\delta_0 Q_+-iD_1\p_\ta Q_+ -\tfrac{1}{2!}D_2\p_\ta^2 Q_+-RQ_-^{(r)}\nn \\ &
	-i\tfrac{1}{2}\kappa (Q_+-\cH_+)-\gamma (|Q_+|^2+2|Q_-|^2)Q_+,\lab{f9a}\\
	 i\p_tQ_-&=\delta_0 Q_- +iD_1\p_\ta Q_- -\tfrac{1}{2!}D_2\p_\ta^2  Q_- -R^*Q_+^{(r)}\nn \\
	&-i\tfrac{1}{2}\kappa (Q_--\cH_-e^{i\ep t})-\gamma (|Q_-|^2+2|Q_+|^2)Q_-, \lab{f9b}\\
	 i\p_tQ_+^{(r)}&=\delta_0 Q_+^{(r)}+iD_1\p_\ta Q_+^{(r)} -
	\tfrac{1}{2}D_2\p_\ta^2 Q_+^{(r)}-RQ_-\nn \\ &
	-i\tfrac{1}{2}\kappa (Q_+^{(r)}-\cH_+)-
	\gamma (|Q_+^{(r)}|^2+2|Q_-^{(r)}|^2)Q_+^{(r)},\lab{f10a}\\
	 i\p_tQ_-^{(r)}&=\delta_0 Q_-^{(r)} -iD_1\p_\ta Q_-^{(r)} -
	\tfrac{1}{2}D_2\p_\ta^2  Q_-^{(r)} -R^*Q_+\nn \\
	&-i\tfrac{1}{2}\kappa (Q_-^{(r)}-\cH_-e^{i\ep t})-\gamma (|Q_-^{(r)}|^2+2|Q_+^{(r)}|^2)Q_-^{(r)}, \lab{f10b}
	\end{align}
\end{subequations}
(see  ~\cite{osac} for the first-principle derivation using the present scaling and notations).
There are several  parameters first used in Equation~\bref{f9} and hence requiring definitions~\cite{osac}.
$\kappa$ is the loaded linewidth parameter. $\gamma$ is the nonlinear coefficients measured,
such that the units of $\gamma/2\pi$ are Hz/W and of $\gamma |Q_\pm|^2/2\pi$ are Hz.
$\cH_\pm$ are the pump parameters, $\cH_\pm^2=\eta\cF\cW_\pm/\pi$. Here, $\eta<1$ is the coupling coefficient,
$\cF=D_1/\kappa$ is the resonator finesse  and $\cW_\pm$ are the on-chip powers of the two lasers. $R$ is the backscattering coefficient,
which is taken equal for all the modes (see  \cite{osac} for a more general treatment).

\subsection{Redefining the Envelope Functions and Eliminating the Repetition-Rate Terms}
\label{H-F_section}
Equations~\bref{f9} contain the $D_1$ terms with the opposite signs, therefore the transformation into the rotating reference frame could remove the repetition-rate term, for example, from the plus equations, but then the repetition rate would simply double in the minus equations. At the same time, $D_1$ is 
the strongly dominant frequency scale in the problem. Typically, it would be 
10 GHz and up to 1 THz, while all the other terms, if taken for the respective range of resonators, 
would vary from 100 kHz to 100 MHz. This indicates that,  if the multiples of  $D_1$ are traced in the  nonlinear terms, then these terms would be oscillating with the fastest frequency in the model and could be disregarded~\cite{osac,cole,lobanov}. 

To  reveal the frequency scales associated with $D_1$, 
we replace the $Q_\mu^{\pm}$ mode amplitudes in Equations~\bref{f3} with
\be
 Q_\mu^\pm(t)=\psi_\mu^\pm(t) e^{-i\mu D_1 t}, 
 \lab{f12}
 \ee
 so that the electric field in Equation~\bref{f1} is  replaced with
\be
\cE = \Big(
e^{iM\vta-i\om_+ t}\sum_{\mu}\psi^+_\mu(t) e^{i\mu\big(\vta-D_1t\big)}+
e^{iM\vta+i\om_+ t}\sum_{\mu}\psi^{-*}_\mu(t) e^{i\mu\big(\vta+D_1t\big)}\Big)+c.c.~.
\lab{f12b}
\ee

Two new sets of the envelope functions 
\be
\psi_{\pm}=\sum_\mu \psi_\mu^{\pm}(t)e^{\pm i\mu\vta},~ 
\psi_{\pm}^{(r)}=\sum_\mu \psi_\mu^{\pm}(t)e^{\mp i\mu\vta},
\lab{f14}
\ee
play a pivotal role in the theory of the microresonators with the 
counter-propagating waves~\cite{osac}. 
Taking the discrete Fourier transforms of $\psi_\pm$  allows   reconstructing  $\cE$ (cf.   
Equations \bref{f1},~\bref{f12} and \bref{f12b}). Unlike the equations for $Q_\pm$, the ones for $\psi_\pm$ 
do not contain the $D_1$ terms  and replace the system for 
four envelopes, $Q_\pm$, $Q_\pm^{(r)}$, with the one for two,
\begin{subequations}
	\lab{f15}
	\begin{align}
	 i\p_t\psi_+&=(\delta_0-2 g_-)\psi_+-\tfrac{1}{2}D_2\p_\vta^2 \psi_+-R\psi_-^{(r)}-\gamma |\psi_+|^2\psi_+ -i\tfrac{1}{2}\kappa (\psi_+-\cH_+),\lab{f15a}\\
	 i\p_t\psi_-^{(r)}&=(\delta_0-2g_+)\psi_-^{(r)}-\tfrac{1}{2}D_2\p_\vta^2 \psi_-^{(r)}-R^*\psi_+-\gamma |\psi_-^{(r)}|^2\psi_-^{(r)}
	-i\tfrac{1}{2}\kappa (\psi_-^{(r)}-\cH_-e^{i\ep t}), \lab{f15b}\\
	g_\pm&=\gamma \int_0^{2\pi}
	|\psi_\pm(t,\vta)|^2 \frac{d\vta}{2\pi}=\gamma \int_0^{2\pi}
	|\psi_\pm^{(r)}(t,\vta)|^2 \frac{d\vta}{2\pi}=\gamma\sum_\mu|\psi_\mu^{\pm}|^2.
	\lab{f15c}
	\end{align}
\end{subequations}

The above equations could be supplemented
with equations for $\psi_+^{(r)}$, $\psi_-$, but this time they  are left as an independent pair. 
With the fabrication techniques rapidly improving,  the surface roughness losses, as well as the associated backscattering, can be reduced to such levels   that the frequency scale of $R$ drops below 
the dispersion and nonlinearity induced effects so that we set $R=0$ in what follows. Including $R\ne 0$
represents a separate problem~\cite{lobanov}.

The notable consequence  of the elimination of the $D_1$ oscillations from the dynamics is that  the nonlinear coupling, i.e.,
the cross-phase modulation (XPM) between the counter-propagating waves, 
has lost its sensitivity to the phases of the individual modes~\cite{osac,cole,lobanov} (see Equation~\bref{f15c}). Now, the XPM is simply expressed by the nonlinear shifts of the detuning, 
i.e., by the $2g_+$ and $2g_-$ terms. 

In what follows, we remove the exponential term from Equation~\bref{f15b} 
by applying an obvious substitution, $\psi_-^{(r)}\to\psi_-^{(r)} e^{i\ep t}$, omit the '$(r)$' superscript, and  deal with
\bsub
\begin{align}
	& i\p_t\psi_+=(\delta_+-2g_-)\psi_+-\tfrac{1}{2}D_{2}\p_\ta^2\psi_+
	-\gamma|\psi_+|^2\psi_+-i\tfrac{1}{2}\kappa(\psi_+-\cH_+),
	\lab{ne1a}
	\\ 
	& i\partial_{t}\psi_{-}=(\delta_{-}-2g_+)\psi_{-}-\tfrac{1}{2}D_{2}\partial_{\theta}^{2}\psi_{-}
	-\gamma|\psi_{-}|^{2}\psi_{-}-i\tfrac{1}{2}\kappa(\psi_{-}-{\cal H_{-}}),
	\lab{ne1b}
\end{align}
\lab{ne1}
\esub
where
\be
\delta_{+}=\delta_0,~~\delta_{-}=\delta_0+\ep.
\ee

\section{Single-Mode, Single-Soliton  and  Soliton-Crystal States  and Their Role in the 
	Soliton-Blockade Effect}\label{sec3}

The single-mode, $\mu=0$, regime of the resonator operation in the plus and minus fields is called here the cw-cw state (where cw stands for {\em continuous wave}). The respective modal amplitudes can be expressed as 
\be\psi_\pm^{(cw)}=\frac{-i\tfrac{1}{2}\kappa\cH_\pm}
{\delta_\pm-2g_\mp-g_\pm-i\tfrac{1}{2}\kappa},\ee 
where $g_\pm$ solve the coupled real algebraic equations
\bsub
\begin{align}
&g_++\frac{4g_+}{\kappa^2}\big([\delta_+-2g_-]-g_+\big)^2=\gamma\cH_+^2,
\\
&
g_-+\frac{4g_-}{\kappa^2}\big([\delta_--2g_+]-g_-\big)^2=\gamma\cH_-^2,
\\
&
g_\pm=\gamma |\psi_\pm^{(cw)}|^2.
\end{align}
\esub

Apart from the cw-cw states, there are also the 
soliton-cw,  cw-soliton and soliton-soliton states, where the first and second words in the solution classification characterize  the field in the  $\psi_{+}$ and $\psi_{-}$ 
components, respectively~\cite{ol}.

Figure~\ref{ff1}a shows how the  cw-cw  solution varies with $\delta_0$ 
for a set of the fixed values of the frequency offset parameter, $\ep=\omega_+ -\omega_-$. 
Large  $|\ep|$  separate and make quasi-independent 
the  resonance structures in the two components along the $\delta_0$ axis, 
i.e., if the $g_+$ vs.  $\delta$ plot has the nonlinearity tilted resonance 
originating at $\delta_0=0$,  then $g_-$ has the similar resonance starting at $\delta_0\approx-\ep$. 
Varying $\ep$  from the relatively large negative to the large positive values drags the $g_-$ resonance across the effectively immobile $g_+$ one (see \mbox{Figure \ref{ff1}a}). Two resonances overlap and interact strongly for $|\ep|/\kappa\lesssim 1$. For $\ep=0$ and $\cH_+=\cH_-$, the model 
becomes symmetric and, therefore, its symmetric solution (see the black lines in the $\ep=0$ panels 
of Figure \ref{ff1}a) goes through the symmetry breaking bifurcation~\cite{ol,chirality,pascal2}. 

The soliton-cw states are solutions of the coupled differential-algebraic system
\begin{subequations}
	\label{so}
	\begin{align}
	 &(\delta_0-2 g_-)\psi_+-\tfrac{1}{2}D_2\p_\ta^2 \psi_+-\gamma |\psi_+|^2\psi_+ -i\tfrac{1}{2}\kappa (\psi_+-\cH_+)=0,\lab{s1}\\
	 &(\delta_0+\ep-2g_+)\psi_- -\gamma |\psi_-|^2\psi_-
	-i\tfrac{1}{2}\kappa (\psi_- -\cH_-)=0,\lab{s2}\\
	 &g_+=\gamma \int_0^{2\pi}
	|\psi_+(\ta)|^2 \frac{d\ta}{2\pi},~~ g_-=\gamma
	|\psi_-|^2. \lab{s3}
	\end{align}
\end{subequations}
If $|\ep|/\kappa$ is  large, then 
Equation~\bref{s1} becomes quasi-independent from Equation~\bref{s2}, implying that the system operates in the limit where it is approximately reduced to the Lugiato--Lefever model 
having the bright soliton solutions 
around the bistability interval for $\delta>0$~\cite{rev3}. However, when $|\ep|$ 
becomes the order of the linewidth,  the two equations start interacting strongly, and the low power branch of the cw-state in the plus-field 
becomes distorted by the growth of the second resonance peak, 
see the $\ep<0$ cases in the top row \mbox{in Figure~\ref{ff1}. }

This interaction also creates the narrow parameter range, 
where the low power branch of the plus field ceases to exist (see the yellow interval in Figure~\ref{ff1}b). Around and substantially beyond this range,  
the soliton pulses in the plus-field are not able to exist simply because the system does not offer a low-amplitude background state for the bright soliton to nest on (see 
the grey stripe in Figure~\ref{ff1}b and the respective no-soliton range in \mbox{Figure~\ref{ff1}c). }
This is the essence of the soliton-blockade effect,
when the solitons can be switched on and off by tuning the frequency of the 
counter-propagating field, so that $\ep$ is tuned within the soliton-forbidden interval~\cite{ol}. 
Figure~\ref{ff2} shows  how the soliton-blockade domains are shaped in the  
$(\delta_0,\ep)$ and $(\cH_-,\ep)$ parameter spaces, respectively.
In particular,  the blockade regime is more easily accessible if the minus , i.e., the cw 
component is pumped harder (see Figure~\ref{ff2}b).

Figure~\ref{ff4} shows how the XPM coefficients for the plus, $g_+$, and minus, $g_-$, 
components of the soliton-cw state (see Equation~\bref{s3})  vary with $\ep$  
for the  value of $\delta$ slightly different from the one in Figure~\ref{ff1}c.
Figure~\ref{ff4}a plots the soliton $g_+$ for the one-, four-, and six-soliton states, i.e., for the soliton crystals. The respective $g_-$ values corresponding to the cw-component are shown in Figure~\ref{ff4}b, 
and the soliton crystal profiles are in Figure~\ref{ff5}.
While the more detailed understanding of the bifurcations 
of the soliton crystals~\cite{cr1,cr2} in the bi-directionally pumped geometry goes beyond our present scope.

\begin{figure}[H]
	\includegraphics[width=0.86\linewidth]{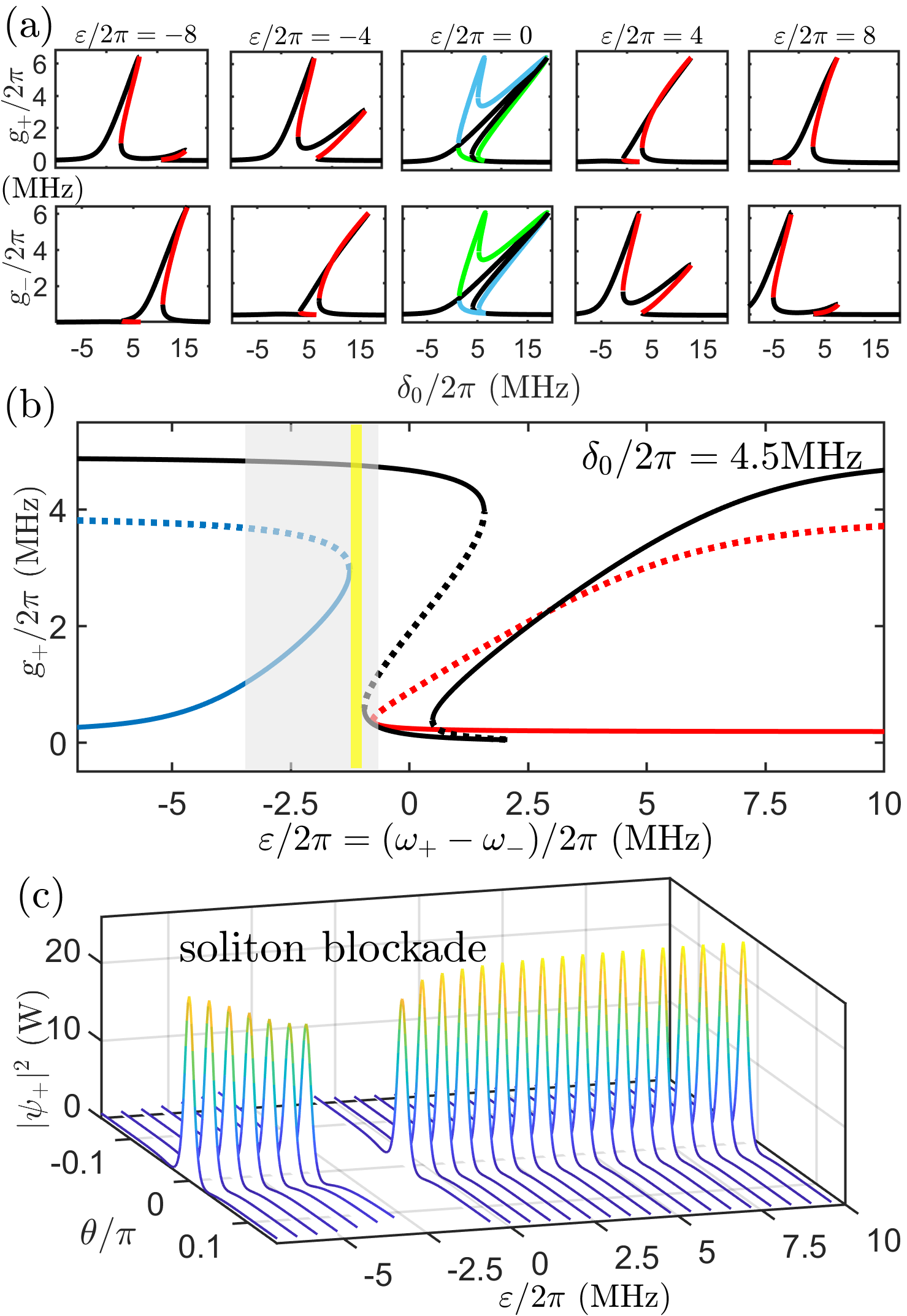}	
	\caption{(\textbf{a}) $g_\pm=\gamma |\psi_\pm^{(cw)}|^2$ for the cw-cw states vs.  
		$\delta_0$ for a set of
		$\ep$. (\textbf{b}) $g_+$ for the cw-cw state vs.  $\ep$ for $\delta_0/2\pi=4.5$ MHz.  (\textbf{c}) Angular profiles of the soliton-cw states vs.  $\ep$. The grey shaded interval in~(\textbf{b}) 
		and the no-solution interval in~(\textbf{c}) is the soliton-blockade interval (see text for further details  and the same and similar datasets published by us in \cite{ol}). Other parameters are $\cH^2_\pm=16$ W, $\kappa/2\pi=1.5$ MHz, 
		$\cF=10^4$, $D_2/2\pi=10$ kHz, $\gamma/2\pi=0.4$ MHz/W, $\eta=0.5$.}
	\lab{ff1}
\end{figure}

 \begin{figure}[H]
 	\includegraphics[width=\linewidth]{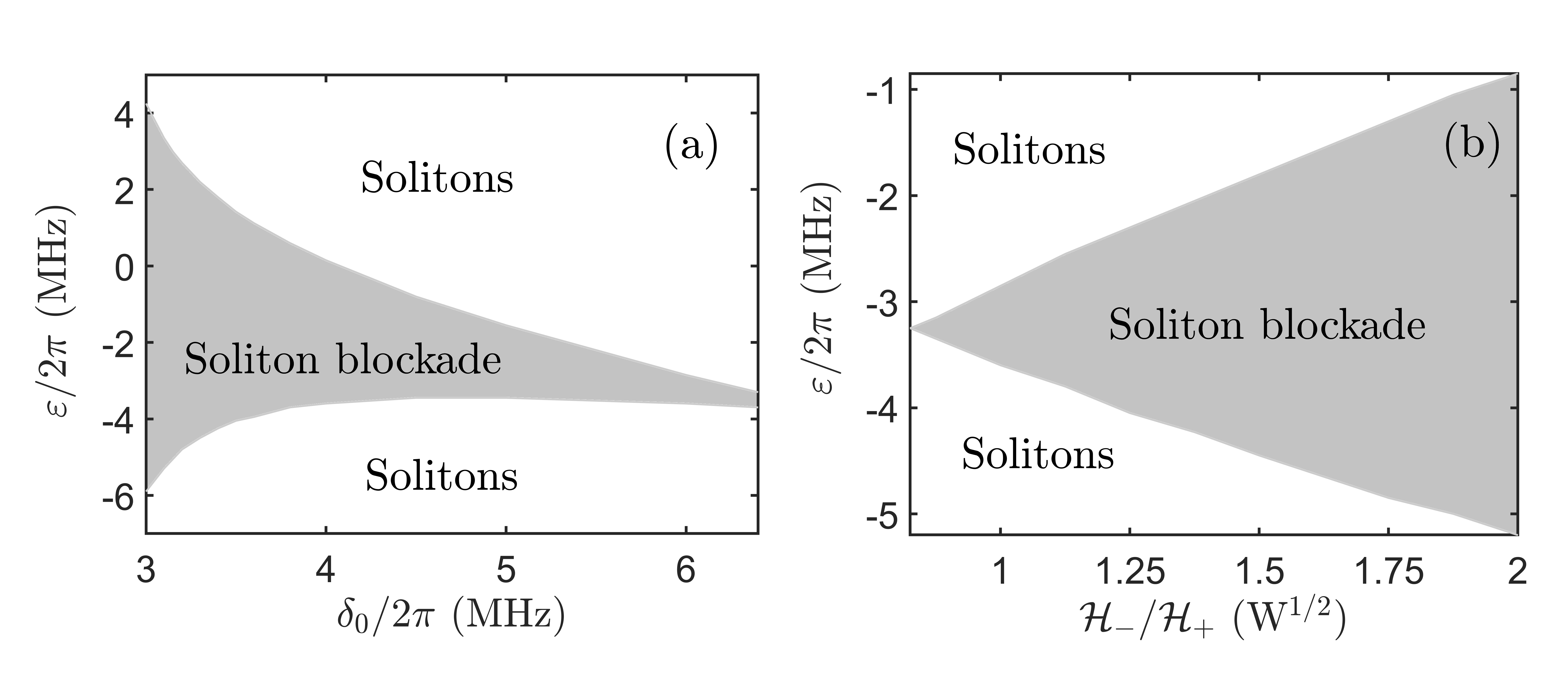}	
 	\caption{The soliton-blockade regions  in the 
 		$(\delta_0,\ep)$  and $(\cH_-/\cH_+,\ep)$ planes. (\textbf{a}) $\cH^2_\pm=16$W; %please confirm if hyphen should be minus
 		(\textbf{b})~$\cH^2_+=16$ W, $\delta_0/2\pi=6$ MHz. 	}
 	\lab{ff2}
 \end{figure} 
\unskip
 \begin{figure}[H]
	\includegraphics[width=0.9\linewidth]{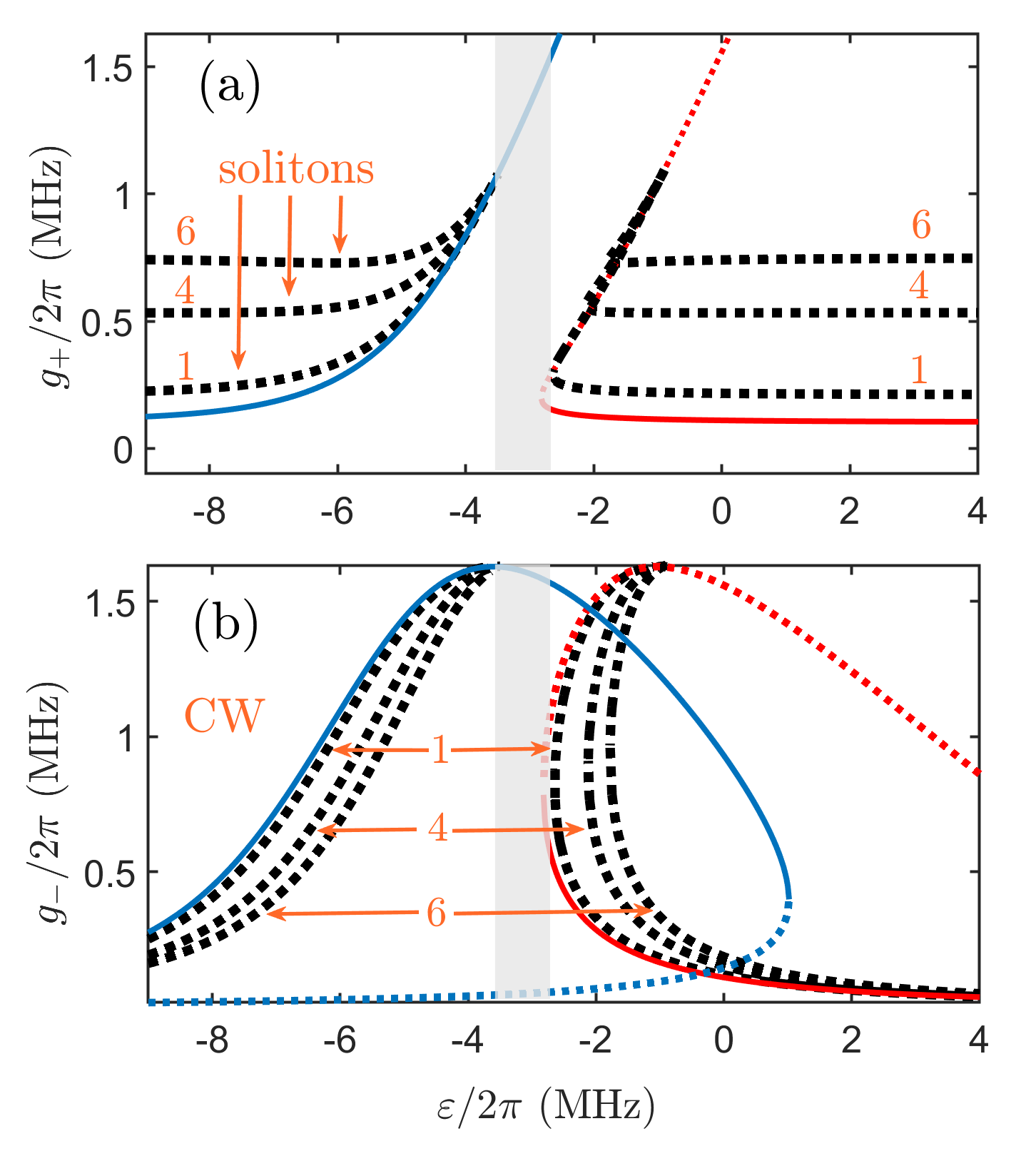}
	\caption{Soliton-cw states for $\delta_0/2\pi=6$ MHz. The other parameters are as in Figure~\ref{ff1}c: %please confirm if hyphen should be minus
		(\textbf{a})   $g_+$ (soliton); and (\textbf{b})   $g_-$ (cw) vs.  $\ep$. The numbers $1$, $4$, and $6$ indicate the soliton-crystal states with the respective  number of the 
		equally spaced solitons (Figure~\ref{ff5}).}
	\lab{ff4}
\end{figure} 

\begin{figure}[H]
	\includegraphics[width=0.8\linewidth]{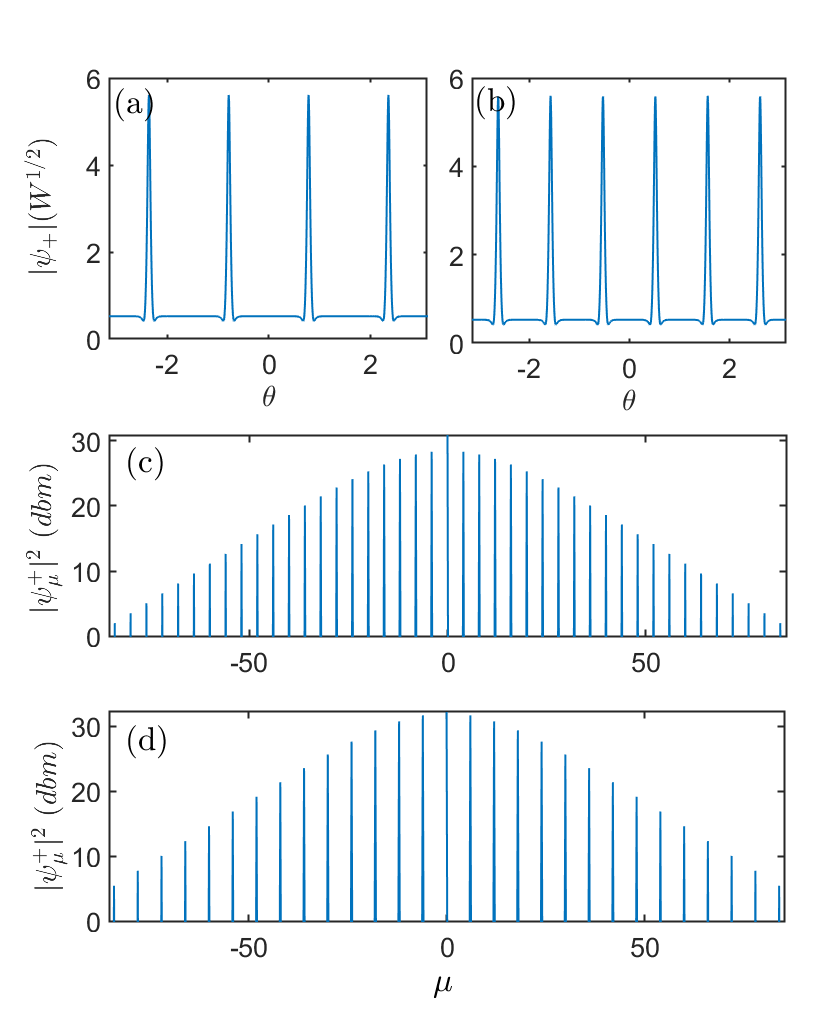}	
	\caption{Soliton crystal states for $\ep/2\pi = 1$ MHz. (\textbf{a},\textbf{b}) are the spatial profiles and (\textbf{c},\textbf{d}) are the respective spectra. The other parameters are as in Figure~\ref{ff4}.}
	\lab{ff5}
\end{figure}

\section{Controlling Solitons by Tuning the Cw-Component: Direct and Reverse Scans}\label{sec4}
We now present the results of the dynamic simulations of Equation~\bref{f15} by initializing them with the cw-cw states and by scanning the frequency of the control (minus) field adiabatically. The adiabatic scans of $\ep$   were  implemented in both forward ($\ep$ goes from negative  to positive) and reverse directions. The outcomes of the two scans are generally different due to the complex and non-symmetric structure of the bistable resonances  (see the cw-cw vs.  $\ep$ plot in Figure~\ref{ff1}b).

First, we choose $\cH_-/\cH_+=1$, and continue to keep $\delta_0$ on the positive side, so that
the plus solitons would always exist in the absence of the interaction with the minus field.
The outcomes of the forward and backward $\ep$-scans are shown in the left and right columns of Figure~\ref{dym1}, respectively. For the forward scan, the plus solitons keep being generated before $\ep/2\pi\simeq -4$ MHz, then disappear in the soliton-blockade region and then reappear for  $\ep/2\pi\gtrsim 7$ MHz. Throughout the blockade interval, the intraresonator field most typically converges to the turbulent state known for the near-bistability operation in the uni-directional setting. For  $\ep/2\pi\gtrsim 2$ MHz, one can also see  that the minus-field starts \mbox{generating solitons. }

In the forward scan, the soliton-blockade interval for the plus field  is wider than
predicted by the time-independent methods in the previous section. 
This is because the eight solitons formed in the minus field provide the value of 
$g_-$ sufficient to keep $\delta_0-2g_-$ outside the plus-soliton existence interval.
Simultaneously, the value of $\delta_0+\ep-2g_+$ is such that the minus field can sustain the solitons. The variations of the net power in either of the fields can lead to different scenarios as $\ep$ is increased. The one that is realized in the dataset shown in Figure~\ref{dym1} is that the downwards fluctuation of $g_+$ takes the minus field
outside the soliton regime and brings it to the low power cw-state. Simultaneously, this transition  restores the soliton regime in the plus field itself. The exit from the blockade regime is also sensitive to the chosen value of $\cH_+/\cH_-$ and the scan parameters. Generally, nonlinear dynamics under conditions when a parameter is swept across the bifurcation points is an area of research attracting attention in optics and beyond (see, e.g.,~\cite{giu,vl} and references therein).

\begin{figure}[H]
	\includegraphics[width=.95\linewidth]{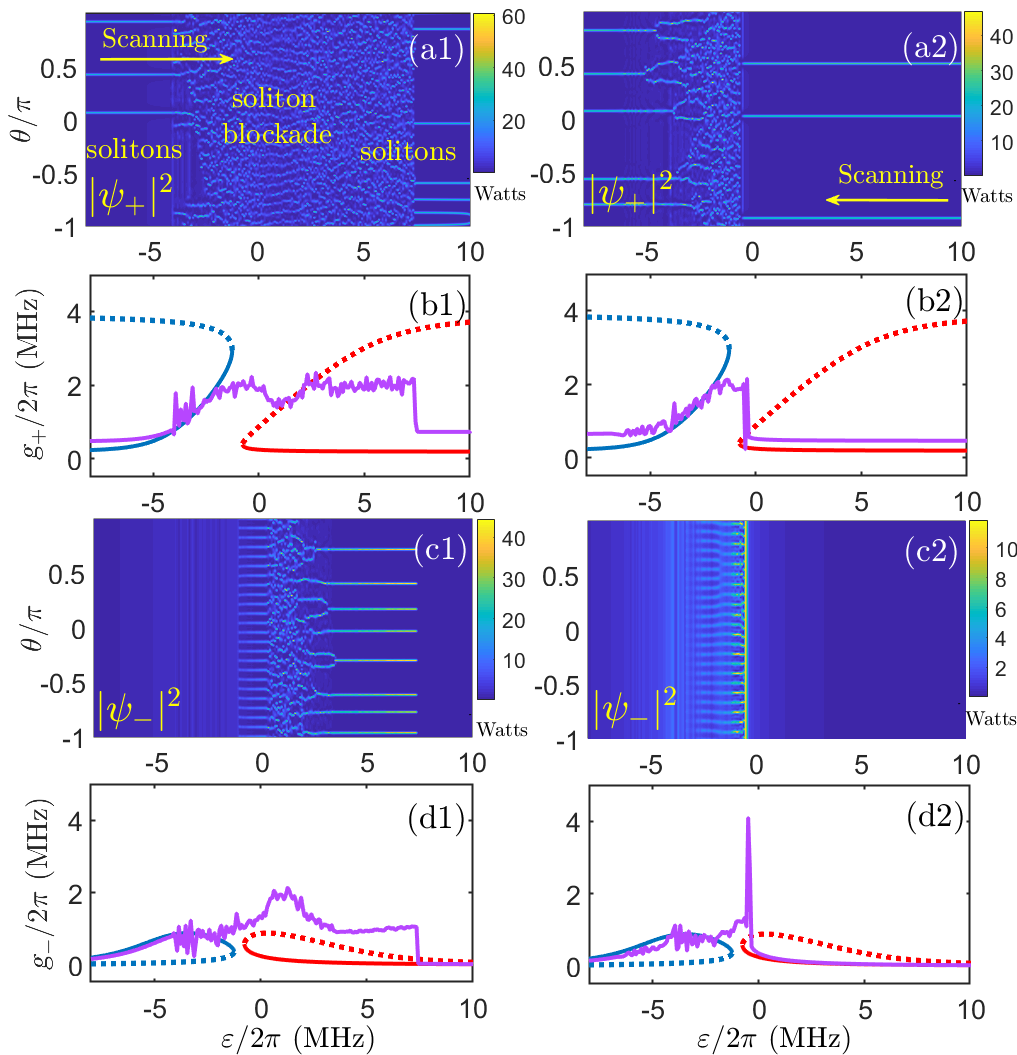}	
	\caption{The soliton blockade in the plus field  as is realized using the 
		forward (\textbf{a1}) and reverse (\textbf{a2}) adiabatic scans of $\ep$, i.e., by tuning $\om_-$; $\delta_0/2\pi=4.5$ MHz, $\cH_+=\cH_-=4$W$^{1/2}$. 
		(\textbf{b1},\textbf{b2}) The magenta line shows  $g_+$ vs.  time. The blue and red lines are $g_+$ for the cw-cw solution. (\textbf{c1},\textbf{c2})   The same as (\textbf{a1},\textbf{a2}) but for the minus field. (\textbf{d1},\textbf{d2}) The same as (\textbf{b1},\textbf{b2}) but for $g_-$.}
	\lab{dym1}
\end{figure} 

The reverse, i.e., $\ep$-positive to $\ep$-negative, scan shows 
the much narrower range of the soliton-blockade in the plus field, 
and no solitons in the minus field (see the right column  
in Figure~\ref{dym1}). This is because the intra-resonator field in the minus component picks and follows the stable low power cw-state until $\ep$ comes to near zero. Therefore, in the reverse scan, the dynamically seen soliton-blockade range is practically the same as the one predicted by the time-independent analysis in the previous section.

We now choose $\cH_-/\cH_+=1/2$, which makes the  soliton-blockade impossible  (see Figure~\ref{ff2}b). The outcomes of the respective forward scan are shown in Figure~\ref{dym4}.  Following our expectations, we see only the stable solitons in the plus field. In the minus field, the periodic pattern is induced in the narrow range of $\delta$s where the low amplitude cw state  is unstable. The uninterrupted solitons seeing in the plus field during the scan are correlated with the continuous existence of the low amplitude cw-state (see the red line in Figure~\ref{dym4}). The narrow and low red bistability peak disturbs the pulse but does not switch the plus field into a persistent non-soliton regime  (cf. Figure~\ref{dym1}). The data in Figures~\ref{dym1} and \ref{dym4} were 
generated while adiabatically tuning $\delta_-/2\pi$ between $-10$ and $30$ MHz over the $1$ ms time interval (the round-trip time $\approx 60$ fs, $D_1/2\pi=15$ GHz).

\begin{figure}[H]
	\includegraphics[width=0.8\linewidth]{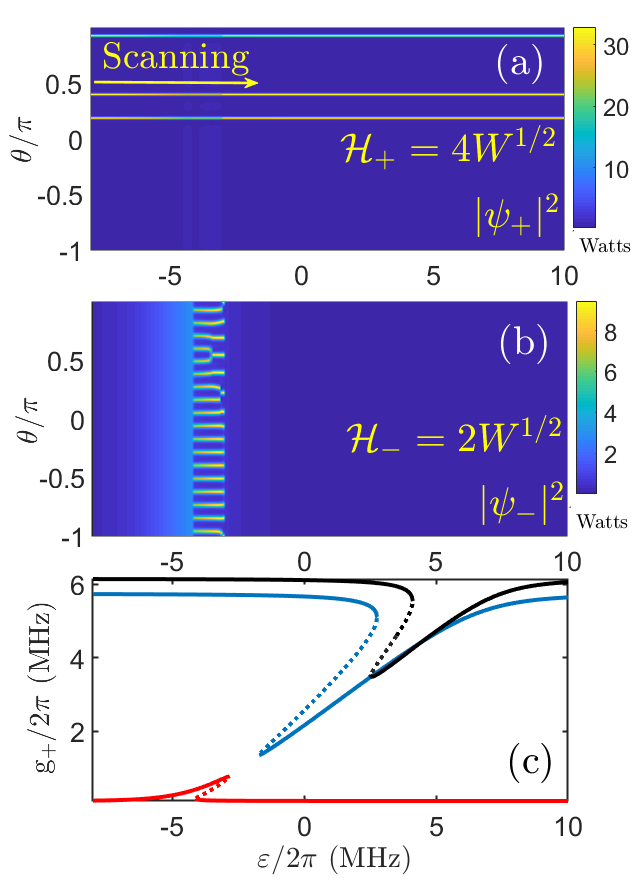}	
	\caption{The forward scan performed outside the soliton-blockade conditions, $\cH_-/\cH_+=1/2$, $\cH_-=2$W$^{1/2}$, $\delta_0/2\pi=6$ MHz (cf.  Figure~\ref{ff2}b): (\textbf{a},\textbf{b})   the plus and minus fields during the scan; and 
		(\textbf{c})    $g_+$  for the cw-cw state over the same interval of $\ep$.}
	\lab{dym4}
\end{figure}

\section{Soliton Stability and Breathers}\label{sec5}
The problem of determining soliton stability  is worthy of some attention, in particular, because  the $g_\pm$ terms in the governing equations represent a certain challenge in this regard.
The soliton stability is determined by computing the eigenvalue of the Jacobian matrix. 
In order to work out the latter, we seek solutions of Equation~\bref{ne1} as $\psi_{+}=A(\theta)+a(\theta,t)$, $\psi_{-}=B(\theta)+b(\theta,t)$, where  
$a,b$ are the small perturbations on top of the soliton or any other spatially inhomogeneous state, $\psi_+=A(\ta)$, $\psi_-=B(\ta)$. 

Truncating the second and higher orders of $a$ and $b$, we are left with 
\begin{align}
	&i\partial_{t}a=\delta_{+}a-\frac{1}{2}D_{2}\partial_{\theta}^{2}a-2\gamma|A|^{2}a-\gamma A^{2}a^{*}-i\frac{1}{2}\kappa a -\frac{\gamma}{\pi} a\int_{0}^{2\pi} |B|^{2} d\theta^\prime \nn\\
	&-\frac{\gamma}{\pi} A\int_{0}^{2\pi}Bb^{*} d\theta^\prime-\frac{\gamma}{\pi} A\int_{0}^{2\pi}B^{*}b d\theta^\prime,   \nn\\
	&i\partial_{t}a^{*}=-\delta_{+}a^{*}+\frac{1}{2}D_{2}\partial_{\theta}^{2}a^{*}+2\gamma|A|^{2}a^{*}+\gamma A^{2*}a-i\frac{1}{2}\kappa a^{*} +\frac{\gamma}{\pi} a^{*}\int_{0}^{2\pi} |B|^{2} d\theta^\prime \nn\\
	&+\frac{\gamma}{\pi} A^{*}\int_{0}^{2\pi}B^{*}b d\theta^\prime + \frac{\gamma}{\pi} A^{*}\int_{0}^{2\pi}Bb^{*} d\theta^\prime,   \lab{ab}\\
	&i\partial_{t}b=\delta_{-}b-\frac{1}{2}D_{2}\partial_{\theta}^{2}b-2\gamma|B|^{2}b-\gamma B^{2}b^{*}-i\frac{1}{2}\kappa b -\frac{\gamma}{\pi} b\int_{0}^{2\pi} |A|^{2} d\theta^\prime \nn\\
	&-\frac{\gamma}{\pi} B\int_{0}^{2\pi}Aa^{*} d\theta^\prime-\frac{\gamma}{\pi} B\int_{0}^{2\pi}A^{*}a d\theta^\prime,   \nn\\
	&i\partial_{t}b^{*}=-\delta_{-}b^{*}+\frac{1}{2}D_{2}\partial_{\theta}^{2}b^{*}+2\gamma|B|^{2}b^{*}+\gamma B^{2*}b-i\frac{1}{2}\kappa b^{*} +\frac{\gamma}{\pi} b^{*}\int_{0}^{2\pi} |A|^{2} d\theta^\prime \nn\\
	&+\frac{\gamma}{\pi} B^{*}\int_{0}^{2\pi}A^{*}a d\theta^\prime + \frac{\gamma}{\pi} B^{*}\int_{0}^{2\pi}Aa^{*} d\theta^\prime.   \nn
\end{align}

If the integrals are replaced with the sums, then Equation~\bref{ab} becomes
$\p_t \vec\ep=-i\wh M\vec\ep$, 
where $\vec\ep=[a,a^{*},b,b^{*}]^{T}$. The growth rate of the perturbations is introduced 
by setting $\vec\ep \sim e^{\lambda t}$. The instability occurs if the real part of $\lambda$ is positive. 
We   computed the soliton spectrum around the blockade interval and   found  the oscillatory 
instabilities that stabilize away from the interval (see Figure~\ref{jcb}a). When the instabilities are present, they typically lead to the formation of the soliton breather states with different periods  (see Figure~\ref{jcb}b,c).

\begin{figure}[H]
	\includegraphics[width=0.75\linewidth]{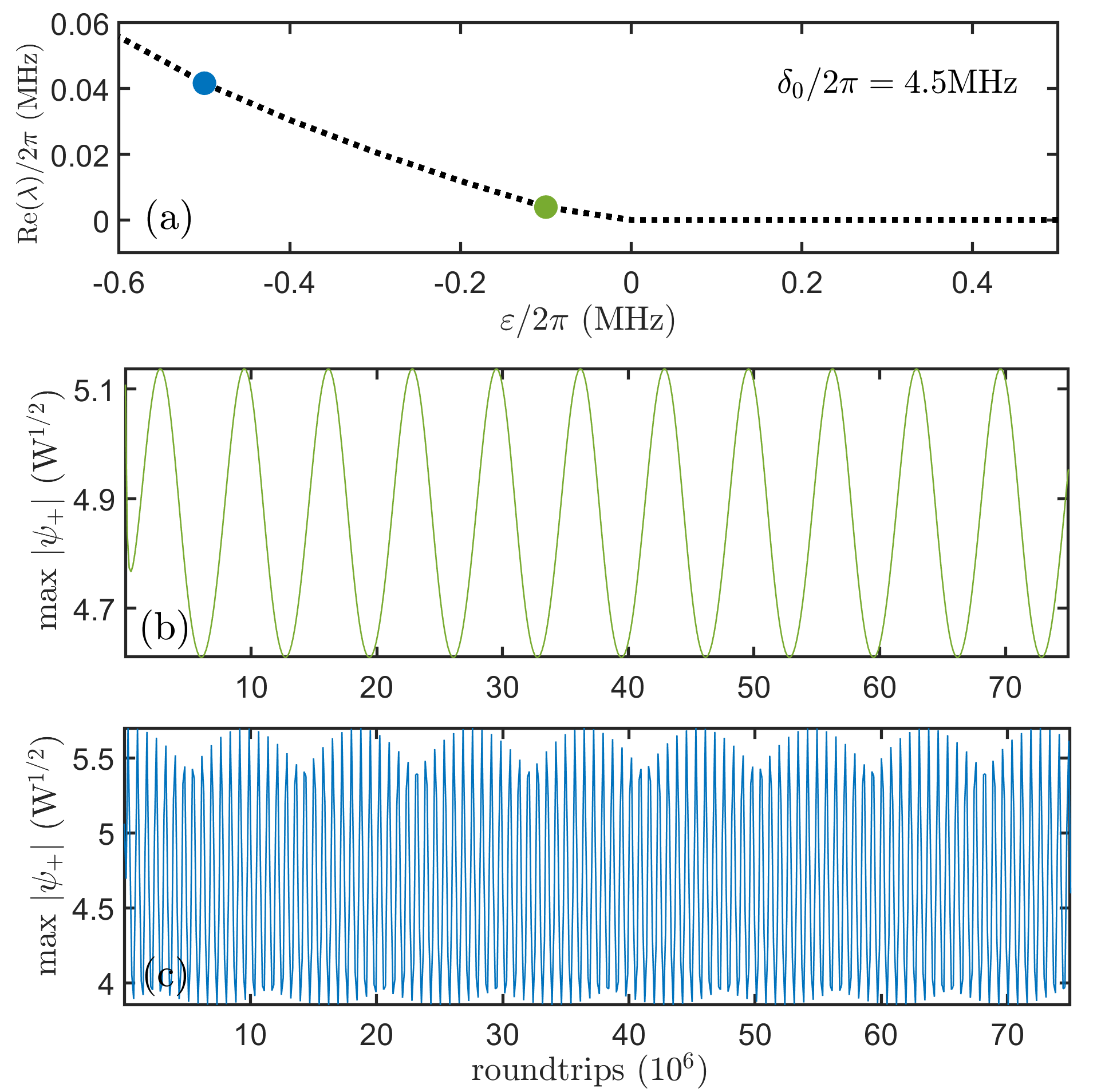}	
	\caption{(\textbf{a}) The maximal growth rate of the soliton instability found on the right from the blockade interval in Figure~\ref{ff1}c. (\textbf{b},\textbf{c}) The soliton breather states found for the green and blue dots in~(\textbf{a}), respectively. }%please confirm if hyphen should be minus
	\lab{jcb}
\end{figure}

\section{Summary}\label{sec6}
We   further investigated  the soliton-blockade effect in the bidirectional microring resonators. In particular, we presented the simulation results on the soliton control by swiping the frequency of the counter-propagating wave in the forward and backward directions  and 
found that the latter provides a much narrower blockade interval and a wider range of the soliton existence.
Furthermore, we   demonstrated that the blockade interval can be expanded by increasing the power of the control field  and that the soliton crystals and  breathers exist on the both sides of it.

\vspace{6pt} 
\authorcontributions{Data curation, Z.F.; Writing---review and editing, D.S.}

\funding{Russian Science Foundation grant number 17-12-01413-$\Pi$;  EU Horizon 2020 Framework Programme (812818, MICROCOMB).}

%\informedconsent{\hl{Not applicable}}%Any research article describing a study involving humans should contain this statement. Please add ``Informed consent was obtained from all subjects involved in the study.'' OR ``Patient consent was waived due to REASON (please provide a detailed justification).'' OR ``Not applicable'' for studies not involving humans. You might also choose to exclude this statement if the study did not involve humans.

%Written informed consent for publication must be obtained from participating patients who can be identified (including by the patients themselves). Please state ``Written informed consent has been obtained from the patient(s) to publish this paper'' if applicable.

%\dataavailability{\hl{Please read the comments in Latex and add, this part is not necessary.}} %In this section, please provide details regarding where data supporting reported results can be found, including links to publicly archived datasets analyzed or generated during the study. Please refer to suggested Data Availability Statements in section ``MDPI Research Data Policies'' at \url{https://www.mdpi.com/ethics}. You might choose to exclude this statement if the study did not report any data.

\acknowledgments{We acknowledge the illuminating discussions with Magnus Johansson.}

\conflictsofinterest{The authors declare no conflict of interest.}

\end{paracol}
\reftitle{References}

%1vexternalbibliography{yes}
%\bibliography{Thebib}

%\begingroup
%\renewcommand{\section}[2]{}%
%\renewcommand{\chapter}[2]{}% for other classes

\end{document}